\documentclass[prl,showpacs,twocolumn,superscriptaddress]{revtex4}
\usepackage[dvipdfm]{graphicx,color}
\usepackage{amsmath, amssymb}

\begin{document}

\title{Impact of geometrical frustration on charge transport near the Mott transition in pyrochlore (Y$_{1-x}$Cd$_{x}$)$_{2}$Mo$_{2}$O$_{7}$}

\author{S. Iguchi}
\author{Y. Kumano}
\author{K. Ueda}
\author{S. Kumakura}
\affiliation{Department of Applied Physics, University of Tokyo, Tokyo 113-8656, Japan}

\author{Y. Tokura}
\affiliation{Department of Applied Physics, University of Tokyo, Tokyo 113-8656, Japan}
\affiliation{Multiferroics Project, ERATO, Japan Science and Technology Agency (JST), Wako, 351-0198, Japan}
\affiliation{Cross-Correlated Materials Research Group (CMRG), ASI, RIKEN, Wako 351-0198, Japan}

\begin{abstract}

To search for novel spin-charge coupled phenomena in the strongly correlated electron system on a frustration lattice, charge transport and dynamics have been investigated for the hole-doped spin-glass (Mott) insulator (Y$_{1-x}$Cd$_{x}$)$_{2}$Mo$_{2}$O$_{7}$ with pyrochlore structure.
Coincident with the disappearance of spin-glass state by hole doping, a diffusive metallic transport with strongly renormalized electron mass shows up with no long range spin order, which is caused by strong antiferromagnetic spin fluctuation due to the geometrical frustration.

\end{abstract}

\maketitle

Concept of geometrical frustration provides a fertile ground for condensed-matter physics. As such an example,
the impact of frustrated spins on itinerant electron dynamics has recently been of great interest.
The conduction electrons coupled with the local spins via the Hund's-rule coupling mediate the magnetic interaction beyond the nearest-neighbor local spins. Then the conduction electrons moving on the background of highly frustrated and antiferromagnetically interacting spins are expected to show novel phenomena, such as heavy electron\cite{LiV2O4-2, LiV2O4}, spin-chiral metal\cite{NdMoO, PrIrO}, superconductivity\cite{CdRe}, etc. 

Among such frustrated spin-charge coupled systems, transition-metal oxides with pyrochlore structure\cite{Subramanian} have been studied extensively to explore novel phenomena near the Mott transition. The pyrochlore lattice, a three-dimensional version of Kagom\'e lattice, provides a stage of frustration of antiferromagnetically interacting spins, to which the conduction electron of $d$ orbital character can be tied via the on-site Hund's-rule coupling.
 Recently, it was found that $R_{2}$Mo$_{2}$O$_{7}$ ($R$ = Nd-Dy) shows an electronic phase transition to a paramagnetic metal (PM) state by the application of high pressure ($\approx$ 10 GPa), irrespective of the starting ground state at ambient pressure, either ferromagnetic metal or spin-glass insulator\cite{RMO-press}. This PM phase is characterized as the nearly temperature-independent resistivity with a slightly larger value than the Ioffe-Regel limit.
The discovery of the diffusive metallic phase in the pyrochlore Mo system stimulated the theoretical investigations on the geometrically spin frustrated Mott transition system \cite{Motome-PRL, Kumar}, and the emergence of the PM phase near the ferromagnetic phase has been understood as the consequence of the competition between ferromagnetic double-exchange and antiferromagnetic superexchange interactions with geometrical frustration\cite{Motome-PRL, Kumar, non-fermi}.
In more general, charge diffusive transport is considered to be generic in the spin-charge coupled system near the Mott transition with geometrical frustration.
Here, we have found that the band-filling control, as another procedure to enhance the charge dynamics in the correlated electron system\cite{RMP}, can realize the similar diffusive PM state even at ambient pressure; the hole doping in Y$_{2}$Mo$_{2}$O$_{7}$ with the insulating spin-glass ground state drives the Mott transition similarly to the case of the pressure-induced transition.
We have found anomalous enhancement of the effective electron mass at around a phase transition point between a spin-glass metallic and a paramagnetic diffusive metallic state.

Pyrochlore-type molybdates $R_{2}$Mo$_{2}$O$_{7}$ are known to show a typical bandwidth-control Mott transition by changing the rare-earth-metal ions from ferromagnetic metal ($R$ = Nd-Eu) to spin-glass insulator ($R$ = Gd-Lu, Y)\cite{R2Mo2O7-phase}.
The sublattice of the Mo ions surrounded by 6 oxygen ions is composed of the corner-shared tetragons\cite{Subramanian}. The MoO$_{6}$ octahedra are compressively distorted along each $\{111\}$ direction, which splits the $t_{2g}$ manifold in the Mo 4$d$ state to the lower $a_{1g}$ singlet and the higher $e_{g}'$ doublet. One of the two electrons in the tetravalent Mo ion in $R_{2}$Mo$_{2}$O$_{7}$ occupies the $a_{1g}$ level as a localized spin, while the other in the $e_{g}'$ as a conduction/localized electron whose spin is aligned parallel to the $a_{1g}$ spin by the strong Hund's-rule coupling.

Polycrystalline samples of (Y$_{1-x}$Cd$_{x}$)$_{2}$Mo$_{2}$O$_{7}$ (YCMO) were prepared by a solid state reaction with the stoichiometric mixtures of powders of Y$_{2}$O$_{3}$, CdO, MoO$_{3}$, and Mo. Partial substitution of Y$^{3+}$ with Cd$^{2+}$ decreases the band filling (or dopes holes) in the $e_{g}'$ conduction electron band. A high-pressure synthesis method was employed in the condition of 4-8 GPa and $1200{}^\circ\mathrm{C}$ to avoid the evaporation of volatile cadmium oxides and to obtain the dense hard pellet whose transport and optical properties could be investigated. 
Powder X-ray diffraction measurements indicated that all the samples were of single phase. 
The resistivity was measured by a 4-probe method, and the magnetization and the heat capacity were measured with a SQUID magnetometer (MPMS, Quantum Design) and by a relaxation method (PPMS, Quantum Design), respectively.  
Optical conductivity spectra were calculated by the Kramers-Kronig transformation of the reflectivity data of 0.1 - 40 eV, with use of the UVSOR facility in the Institute for Molecular Science in the photon energy above 4 eV.

\begin{figure}
\includegraphics*[width=8.0cm]{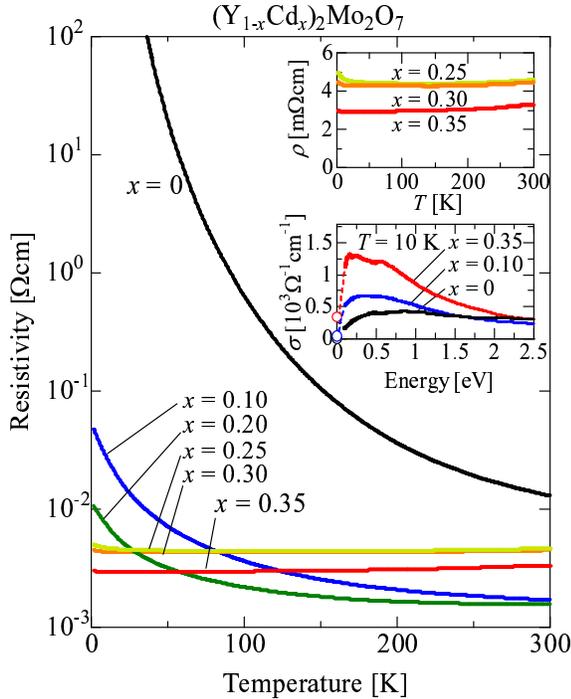}
\caption{(Color online) Temperature dependence of resistivity for (Y$_{1-x}$Cd$_{x}$)$_{2}$Mo$_{2}$O$_{7}$ (YCMO). The inset shows (upper) the magnified views of the resistivity for $x=0.25, 0.30, 0.35$ on a linear scale and (lower) the optical conductivity for YCMO $x=0, 0.1$, and 0.35 in comparison with the dc conductivity (open circles at 0 eV) at 10 K. Dotted lines are merely the guide to the eyes.}
\label{fig1}
\end{figure}

The temperature dependence of resistivity in YCMO is shown in Fig. 1.
The resistivity for $x=0$ is insulating as is known\cite{Y2Mo2O7}, while the 10\% hole doping ($x=0.10$) makes the charge transport metallic. 
Further doping reduces the resistivity monotonically, although it is slightly larger than the Ioffe-Regel limit of the system ($\sim$ 0.5 m$\Omega$ cm \cite{RMO-press}).
Among the metallic specimens the resistivities in $x \geq 0.25$ show almost no temperature-dependence from 300 K to 2 K apart from a slight upturn in the low temperature region ($\leq$ 30 K, see the inset of Fig. 1). The deviation of the resistivity in the whole temperature region is within 10\%, while the resistivity value itself decreases with doping content $x$.
These anomalous features are quite similar to the diffusive transport observed for the PM phase induced by application of high pressures in undoped $R_{2}$Mo$_{2}$O$_{7}$ (single crystals) \cite{RMO-press}.
The magnetoresistance at 9 T in the diffusive metallic specimens is less than 1\% at any temperature (not shown here).
The diffusive metal state observed also in this hole-doped system is considered as the intrinsic property of the PM phase near the Mott critical point in the pyrochlore-type molybdates.
The optical conductivity spectra $\sigma(\omega)$ for YCMO ($x=$0, 0.10, and 0.35) were measured at 10 K, shown in the lower inset of Fig. 1. Y$_{2}$Mo$_{2}$O$_{7}$ ($x=0$) shows a small Mott gap ($\leq$ 0.1 eV) which is consistent with the data of the melt-grown polycrystalline sample\cite{Y2Mo2O7} and the other insulating single crystals of $R_{2}$Mo$_{2}$O$_{7}$ ($R=$Gd-Ho)\cite{RMO-istvan}.  The $\sigma (\omega)$ is enhanced in magnitude with increasing of $x$ in the $d$-$d$ transition region ($\leq 2$ eV). The disappearance of the gap in $x \geq 0.1$, i.e. the remaining optical conductivity $\sigma (\omega \to 0)$, corresponds to that of the dc conductivity at 10 K represented by the open circles in the inset, while even the high-doped ($x=0.35$) sample appears to show the pseudo-gap like feature below 0.2 eV in the $\sigma(\omega)$ spectrum.


\begin{figure}
\includegraphics[width=9.0cm]{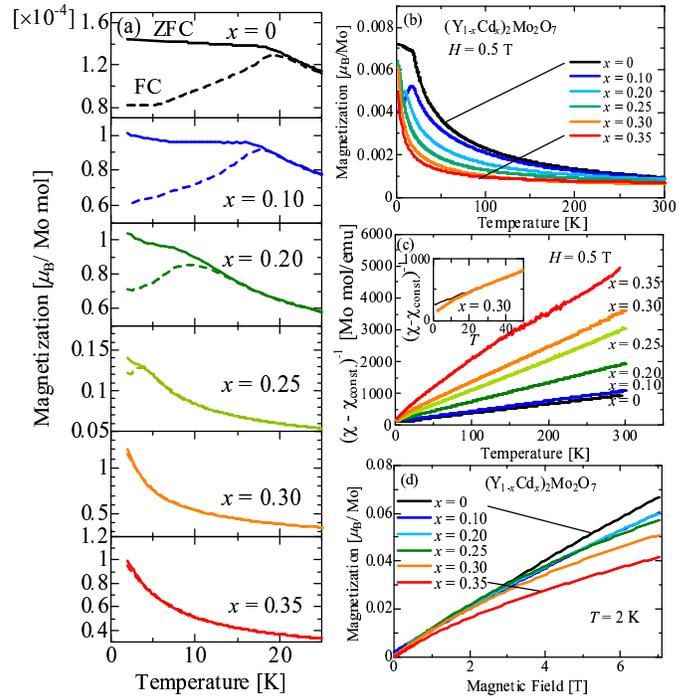}
\caption{(Color online) Temperature dependence of (a) magnetization in the process of field cool (FC) and zero-field cool (ZFC) at 0.01 T, (b) magnetization at 0.5 T, (c) inverse susceptibility subtracted by constant term ($\chi_{const.}$) as determined by the best-fit result to the Curie-Weiss law for (Y$_{1-x}$Cd$_{x}$)$_{2}$Mo$_{2}$O$_{7}$.  Inset in (c) shows the magnified view of inverse susceptibility for $x=0.30$ at low temperatures. (d) Magnetic field dependence of magnetization at 2 K.}
\label{fig2}
\end{figure}

The magnetic properties are shown in Fig. 2; the temperature dependence of (a) field-cooled (FC) and zero-field-cooled (ZFC) magnetization at 0.01 T, (b) magnetization at 0.5 T, (c) inverse susceptibility, and (d) magnetic field dependence of magnetization at 2 K.
In Fig. 2(a), the difference between the FC- and ZFC-magnetizations is clearly discerned for $x\leq 0.25$ indicating the subsistence of the spin-glass state, while no longer for $x\geq 0.30$.
The spin-glass transition temperature decreases monotonically with increasing hole doping and finally disappears at $x=0.30$ without any sign of other magnetic transitions. 
Note that the disappearance of spin-glass state is not concurrent with the insulator-to-metal transition. YCMO ($x=0.25$) still shows the spin-glass feature at 3.5 K, while the transition from insulator to metal occurs at below $x=0.1$.
Thus, the PM state realized in the hole doped YCMO ($x \gtrsim 0.3 $) is distinctly above the insulator-to-metal transition point ($x_{c} \sim 0.1$), and is expected to exist widely in $x\geq 0.30$. This is analogous to the case of the pressure-induced transitions in $R_{2}$Mo$_{2}$O$_{7}$\cite{RMO-press}; the PM phase spreads at higher pressures well above the metal-insulator critical pressure.

The magnetization decreases monotonically with increasing $x$ [see Figs. 2(b) and (d)], and the temperature dependence of magnetization seems to follow the Curie-Weiss law. Figure 2(b) shows the best-fit results to the Curie-Weiss law with a $T$-independent constant term expressed as
\[
\chi{}(T) = \frac{C\mu_{eff}^{2}}{T-\Theta_{CW}} + \chi{}_{const.},
\]
where $C$, $\mu_{eff}$, $\Theta_{\rm{CW}}$, and $\chi_{const.}$ represent the Curie constant, effective moment, Curie-Weiss temperature, and constant term, respectively.
The inverse susceptibility data, subtracted by $\chi_{const.}$, are well represented by the law with finite effective moments and negative Weiss-temperatures which are comparable with those of Y$_{2}$Mo$_{2}$O$_{7}$ [see Fig. 4(b)]. The $|\Theta_{\rm{CW}}|$ in Y$_{2}$Mo$_{2}$O$_{7}$ is known to be reduced below room temperature, while in high temperature region it shows $\sim 200$ K according to the literature\cite{YMO-glass-Raju, YMO-glass-Greedan}.
The temperature dependence of the magnetization of the paramagnetic YCMO($x=0.30$) deviates from the Curie-Weiss law in the lower temperature region ($\leq 30$ K) as shown in the inset of Fig. 2(c), implying the possible enhancement of $\chi_{const.}$.
The decrease in the magnetization is seen also in the magnetic field dependence of magnetization at 2 K in Fig. 2(d). This is ascribed to the reduction of the electrons on the $e_{g}'$ state whose spins are aligned parallel with the localized $a_{1g}$ spins by the strong Hund's-rule coupling. The almost linear dependence of the magnetization with magnetic field changes into an upward convex curve, implying the reduced antiferromagnetic interaction.
In contrast to the 3$d$ electron heavy fermion system LiV$_{2}$O$_{4}$\cite{LiV2O4-2, LiV2O4}, the magnetic moment of $x=0.30$ seems to remain finite at the lowest temperature. 
This means that the metallic ground state of YCMO ($x \geq 0.30$) is still under antiferromagnetic spin correlation, while no long range order is present perhaps due to the high frustration on the pyrochlore lattice.

\begin{figure}
\includegraphics[width=8.5cm]{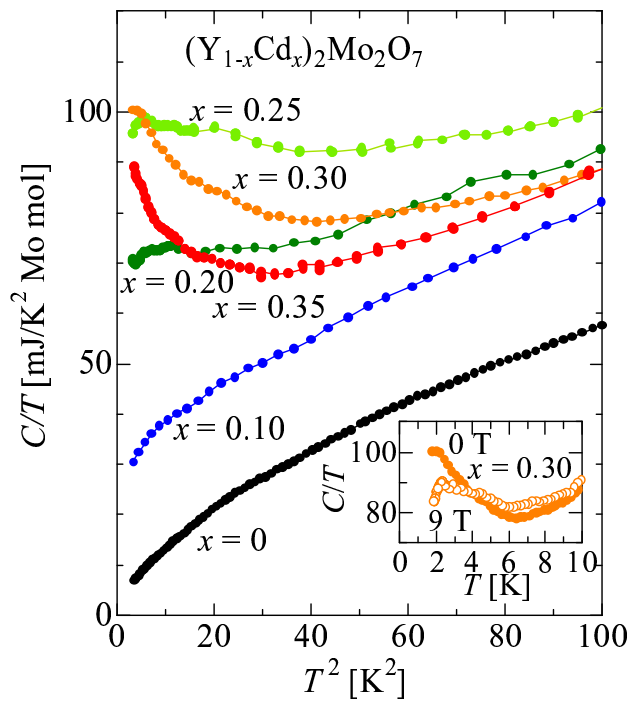}
\caption{(Color online) $T^{2}$ dependence of heat capacity ($C$) divided by temperature ($T$) in (Y$_{1-x}$Cd$_{x}$)$_{2}$Mo$_{2}$O$_{7}$. $C/T$ under the magnetic field of 9 T for $x = 0.30$ is shown in the inset.}
\label{fig3}
\end{figure}

Figure 3 shows the heat capacity ($C$) divided by temperature ($T$) for (Y$_{1-x}$Cd$_{x}$)$_{2}$Mo$_{2}$O$_{7}$.
In the insulating phase, the electronic specific heat coefficient ($\gamma$), defined as the $C/T$ value at the lowest temperature, converges to zero as $T \to 0$ for $x=0$.
The $C/T$ vs. $T^{2}$ curve in $x=0.10$ shifts upward by $\sim$ 25 mJ/K$^{2}$Mo$\cdot{}$mol, indicating the closing of the Mott gap and the evolution of the density of state at the Fermi level in the system. The $\gamma$ value for $x=0.10$ is twice as large as that in the nearly-ferromagnetic (ferromagnetic cluster-glass) metallic phase of (Eu, Y)$_{2}$Mo$_{2}$O$_{7}$ $\sim$ 13 mJ/K$^{2}$Mo$\cdot{}$mol\cite{Hanasaki-Eu}.
The $C/T$ vs. $T^{2}$ curve in the low temperature region qualitatively changes above $x \sim 0.20$, corresponding to the appearance of diffusive metallic feature. Anomalous enhancement of the $C/T$ value at low temperatures is clearly seen for $x=0.30$ which is in the vicinity of the phase boundary between the SGM and PM phase.
The $\gamma$ value for $x=0.30$ becomes as large as 100 mJ/K$^{2}$Mo$\cdot{}$mol, which reaches more than one half of that of the heavy fermion system, LiV$_{2}$O$_{4}$ ($\sim$170 mJ/K$^{2}$V$\cdot{}$mol\cite{LiV2O4-2}). Further increase of doping level tends to suppress $\gamma$, although the anomalous enhancement at the low temperature is still observed.
The suppression of $\gamma$ by the application of magnetic field of 9 T is seen in the inset of Fig. 3, indicating the partial suppression of the spin fluctuation subsisting down to the lowest temperature; this is also supported by the fact that both the enhancement and field-suppression of $\gamma$ become maximal at the phase boundary between SGM and PM [see Fig. 4(d)].


\begin{figure}
\includegraphics[width=8.0cm]{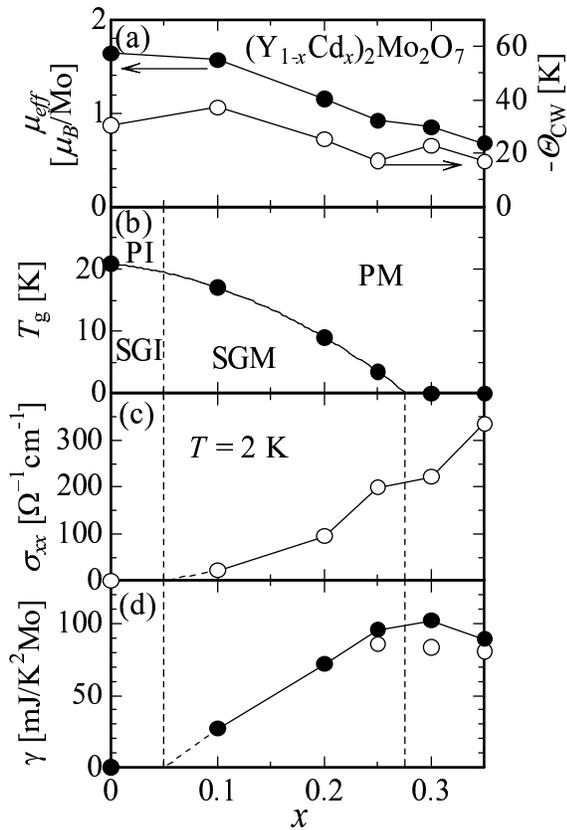}
\caption{The $x$ dependence of (a) the effective magnetic moment ($\mu_{eff}$) and Weiss temperature ($\Theta_{CW}$), (b) magnetic state, (c) dc conductivity at 2 K, and (d) $\gamma$ defined as the $C/T$ at 1.8 K. The PI, PM, SGI, and SGM in (b) stand for paramagnetic insulator, paramagnetic metal, spin-glass insulator, and spin-glass metal, respectively. The open circles in (d) show the $\gamma$ values at 9 T.}
\label{fig4}
\end{figure}

Figure 4 shows the $x$ dependence of (a) the effective magnetic moment ($\mu_{\rm{eff}}$), (b) the magnetic state, (c) the conductivity at 2 K, and (d) the $\gamma$ for YCMO.
The insulator to metal transition occurs between $x=0$ and $x=0.10$ and the diffusive metallic state appears in $x \geq 0.25$, while the spin-glass state still survives up to $x=0.25$ and the comparable magnitude of effective spin moment with that of Y$_{2}$Mo$_{2}$O$_{7}$ is discerned up to above $x=0.3$.
The $\gamma$ increases with $x$ up to the maximum value of $\sim$100 mJ/K$^{2}$Mo$\cdot{}$mol at $x=0.30$, where the spin-glass disappears and the field-induced suppression of $\gamma$ also becomes maximal. Given that the conduction electron density is 0.7 ($x=0.3$) per Mo in the simple parabolic band, the observed $\gamma$ value would give the effective electron mass as large as 75$m_{e}$, $m_{e}$ being the bare electron mass.
Note that if the susceptibility at the lowest temperature were due to the enhancement of Pauli paramagnetism, the calculated Wilson ratio would exceed 20, being unphysical. The origin of the critical mass enhancement on the verge of the PM-to-SGM transition may be the strong antiferromagnetic spin fluctuation assisted by the geometrical frustration effect. This renders the present pyrochlore compounds distinct features from the conventional heavy fermion system and also perhaps the non-Fermi liquid nature as predicted theoretically for the relevant model system\cite{non-fermi}.

In summary, we have investigated the effect of geometrical frustration on the spin-charge coupled phenomena in the hole-doped pyrochlore molybdates (Y$_{1-x}$Cd$_{x}$)$_{2}$Mo$_{2}$O$_{7}$(YCMO) near the Mott transition.
The spin-glass transition temperature is effectively reduced by Cd doping (hole doping), resulting in the paramagnetic state with strong antiferromagnetic spin correlation assisted by the geometrical frustration.
The diffusive paramagnetic metallic (PM) state appears at around $x=0.30$ in YCMO, in close analogy to that under high pressure in $R_{2}$Mo$_{2}$O$_{7}$, suggesting the generic appearance of the PM phase near the Mott transition in this system.
Around the phase boundary between spin-glass metal (SGM) and PM phase, the strong electron mass renormalization was observed.
We can ascribe such an anomalous metal state to the coupling of the conduction electrons with the strong antiferromagnetic spin fluctuation as generated via the geometrical frustration.

We thank Y. Motome and N. Furukawa for helpful discussions. This work was partly supported by Grants-in-Aid for Scientific Research 
(Nos. 22740220
, 20340086
, and 22014003
), MEXT of Japan 
and JSPS through its FIRST Program.


\end{document}